\documentclass[a4paper, english,11pt]{article}
\usepackage{color}
\usepackage{dirtree}
\usepackage[utf8]{inputenc}
\usepackage[T1]{fontenc}
\usepackage{wrapfig}
\usepackage{float}
\usepackage[usenames, dvipsnames]{xcolor}
\usepackage{amsmath}
\usepackage{amssymb}
\usepackage{mathrsfs}
\usepackage{makeidx}
\usepackage{enumitem}
\usepackage{tabularx}
\usepackage{stmaryrd}
\usepackage{mathtools}
\usepackage{bm}
\usepackage{import}
\usepackage[left=2.5cm,right=2.5cm,bottom=3cm]{geometry}
\usepackage{graphicx}
\usepackage{makecell}
\usepackage{fancyhdr} 
\usepackage{comment}
\usepackage{indentfirst}
\usepackage{verbatim}
\usepackage[table]{colortbl}
\usepackage{pgfplotstable}
\usepackage[toc,page]{appendix}
\usepackage{nicematrix}
\usepackage{amsmath,amssymb}
\NiceMatrixOptions{nullify-dots}
\usepackage{svg}
\usepackage{subcaption}
\usepackage[numbers]{natbib}
\usepackage{authblk}
\usepackage{gensymb}
\usepackage{chngcntr}
\usepackage{tikz}
\usetikzlibrary{patterns}
\usetikzlibrary{shapes}
\usetikzlibrary{arrows.meta}
\usetikzlibrary{calc,shadows,matrix}
\usetikzlibrary{decorations.pathreplacing}
\usepackage{indentfirst}
\usepackage{hyperref}
\usepgfplotslibrary{fillbetween}
\usepackage{pgfplots}
\usepgfplotslibrary{patchplots}
\usepgfplotslibrary{fillbetween}
\usepgfplotslibrary{polar} 
\usetikzlibrary{pgfplots.polar}
\usepgfplotslibrary{colormaps}
\newcommand{\imag}{\mathrm{Im}}				%Imaginary part
\title{Improving the vibration isolation efficiency of chiral phononic crystals through viscoelastic damping}
\author[1,2]{L. Mardini}
\author[1]{A. Bergamini}
\author[3,4]{C. Claeys}
\author[2,4]{E. Deckers}
\author[1]{B. Van Damme}
\date{}
\affil[1]	{Empa, Laboratory for Acoustics/Noise Control, Ueberlandstrasse 129, 8600 Dübendorf, Switzerland}

\affil[2]	{Department of Mechanical Engineering campus Diepenbeek, KU Leuven
Wetenschapspark 27, B-3590, Diepenbeek, Belgium}

\affil[3]		{Department of Mechanical Engineering campus Heverlee, KU Leuven, Celestijnenlaan 300, B-3001, Heverlee, Belgium }

\affil[4]		{Flanders Make@KU Leuven, Belgium }
\pgfplotsset{compat=1.18} 
\begin{document}
\maketitle

\section*{Abstract}

Mechanical phononic crystals attenuate vibration transmission in a wide frequency range called the bandgap. Chiral phononic crystals have demonstrated their ability to achieve low-frequency bandgaps while maintaining a high static stiffness. However, their design is non-trivial because of the complex geometry involved. Furthermore, bandgap generation in phononic crystals relies on wave scattering rather than energy dissipation, which leads to important local vibration amplitudes. In this paper, viscoelastic inserts are introduced with two goals: improve the crystal’s tunability in terms of stiffness and bandgap properties, and employ damping to reduce internal resonances.
First, the fundamental effect of viscous damping in phononic crystals is investigated through analytical equivalent models, showing that moderate damping improves the vibration isolation by merging multiple bandgaps.
As a real-world example, highly viscoelastic polyurethane is studied by examining the influence of its frequency-dependent storage modulus and loss factor. Three crystals with different viscoelastic inserts are manufactured and the measured force transmissibility is successfully compared to the numerical models. The results show that the storage modulus correlates strongly with vibration isolation efficiency, defining both the frequency range and the maximum force reduction, whereas a loss modulus broadens the vibration isolation frequency range without negatively affecting its efficiency when it stays below a critical threshold value. 

\section{Introduction}
Vibration isolation is a well-known challenge in numerous fields of engineering. Industrial machinery, buildings, and vehicle cabins require a high level of vibration isolation from their surrounding environment to either protect workers' health~\cite{gafsi_modeling_2017,lee_reduction_2016}, maintain structural integrity~\cite{sheng_development_2022} or ensure passenger comfort~\cite{papaioannou_optimal_2020}.   
Passive linear isolators, resilient elements placed between the vibrating source and the environment, are commonly used due to their low cost and simple design. They attenuate vibrations above their mass-spring resonance frequency~\cite{nilsson_vibro-acoustics_2015} which is dependent on the sprung mass and inherent axial stiffness of the isolator. 
At the resonance frequency, the force transmitted is amplified. By increasing the inherent damping of the isolator, this amplification can be reduced but an increase in the damping in turn decreases the vibration isolation above the resonance frequency~\cite{schiavi_improvement_2018}. 

\begin{figure}
    \centering
     \includegraphics[width=0.5\linewidth]{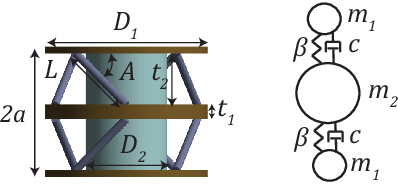} 
    \caption{Schematic representation of a chiral phononic crystal unit cell with equivalent systems of masses, springs and dampers.}
    \label{fig:UCMassSpring}
\end{figure}

More recently, phononic crystals (PCs) have been investigated as vibration isolators, exploiting Bragg bandgaps to attenuate wave propagation over a wide frequency range~\cite{brillouin_wave_1953,deymier_acoustic_2013}. A summary of recent research on the application of PCs for vibration isolation is presented in~\cite{huang_application_2025}. PCs can be represented as a periodic chain of masses and springs as shown in Figure~\ref{fig:UCMassSpring}. 
Due to resonant scattering by the masses, undamped wave propagation is blocked in certain frequency ranges, the so-called bandgaps.
A PC is an efficient vibration isolator within the bandgap, where the 
dynamic force transmission is reduced. 
For both resilient isolators and PCs, low-frequency isolation without increasing the mass can only be achieved by decreasing the support stiffness. 

Chiral phononic crystals (CPCs) respond to this challenge by maintaining the spring stiffness and mass while shifting the longitudinal bandgap to a lower frequency
~\cite{bergamini_tacticity_2019,ding_description_2023}. The enhanced kinematics between the masses of the chain, made possible by its slanted connectors as shown in Fig.~\ref{fig:UCMassSpring}, amplifies their inertia through the addition of rotation to the motion of the disks when the CPC is compressed. 
However, CPCs are complex to optimize. In the case shown in Figure~\ref{fig:UCMassSpring}, the middle disk experiences strong rotational coupling and therefore exhibits higher inertia than the outer disks. The total inertia $m_2$ of the rotating masses can be explicitly calculated based on their static mass $m_1$, moment of inertia, and angle of inclination of the connectors. This type of CPC can therefore be represented by an equivalent two-mass PC, as shown in Figure~\ref{fig:UCMassSpring}. The axial stiffness either stems from the tilted connectors and is dependent on their shape, material properties, inclination angle and connection to the disks~\cite{bergamini_tacticity_2019}, or can be easily tuned by an improved design that decouples the bending of the connectors from the axial stiffness between the masses~\cite{ding_origin_2024}. 

Bandgaps in PCs originate from wave interference which causes amplitude attenuation in transmission, but inevitably introduces energy reflection. Adding damping introduces dissipation, which reduces the amplitude of the reflected waves. Previous works have been investigating the effect of damping on the behavior of PCs. 
Depending on the type of damping considered (structural, viscous, or viscoelastic) an upward or downward frequency shift of the dispersion curves can be observed~\cite{andreassen_analysis_2013,hussein_metadamping_2013,frazier_viscous--viscoelastic_2015}, and the effect of damping on both spatial and temporal attenuation has been assessed~\cite{farzbod_analysis_2011,frazier_viscous--viscoelastic_2015,andreassen_analysis_2013}. A fundamental study by \citet{wang2015visco} showed that introducing viscoelastic elements in phononic crystals alters both the real and imaginary part of the wavenumber. \citet{moiseyenko_material_2011} showed that increasing damping results in smoother curves for the real and imaginary part of the wavenumber and increase spatial attenuation at internal resonances.   
In general, damping in PCs introduces spatial attenuation of waves inside and outside the bandgaps, since the wave number always has a non-zero imaginary part~\cite{farzbod_analysis_2011}. Similarly, for transient excitations, an increase of the viscous~\cite{frazier_viscous--viscoelastic_2015} and structural~\cite{andreassen_analysis_2013} damping leads to faster attenuation of the dynamic response over the full frequency range. 

Despite this multitude of studies, damping is mostly studied in a simplified way, omitting frequency-dependency of the stiffness, or assuming a constant structural or viscous loss factor. Realistic conventional materials used for vibration isolation exhibit a rich damping behavior~\cite{baz_active_2019,rouleau_application_2013}. 
The dynamic modulus in viscoelastic materials under harmonic excitation is complex and frequency-dependent: 
\begin{equation}
    E(\omega) = E'(\omega) + iE''(\omega), \label{complexModulus}
\end{equation}
where $E'$ is the storage modulus, a measure of the stored energy, and $E''$ the loss modulus, a measure of the dissipated energy. The loss factor is defined as $\eta = E''/E'$. The full frequency-dependent material behaviour is typically presented in the form of master curves which change with temperature. 
The relation between storage modulus, loss factor and frequency can be given by material descriptions such as the generalized Maxwell model or fractional derivative Zener model~\cite{baz_active_2019}. Realistic viscoelastic materials have also been investigated in metamaterial research, in particular for locally resonant metamaterials~\cite{lewinska2017attenuation,krushynska_visco-elastic_2016}, showing that the frequency-dependent stiffness can shift the bandgap frequency, and that higher loss factors reduce the imaginary part of the wave number.  In~\cite{zhu_band_2016,zhao_band_2009}, the influence of viscoelastic model parameters on the real part of the wavenumber in phononic crystals is studied, and~\cite{beniwal2026predictive} shows that frequency-dependent material parameters lead to better prediction of the complex dispersion in 1-D PCs. More advanced applications, such as topological~\cite{ma2022complex} and adaptive~\cite{dykstra2022extreme} materials were shown to profit from the inclusion of viscoelastic materials.

In this work, the practical property range of a strongly viscoelastic material, polyurethane foam, is exploited within the design of CPCs. Its strongly frequency-dependent master curves do not allow simplified modeling using constant stiffness and damping. The material is integrated into a practical design of CPCs with a dual goal: to allow easy bandgap tuning based on stiffness variation, and to dissipate elastic energy without negatively affecting the crystal's vibration isolation capacity. Similar to the designs presented in \cite{ding_origin_2024,qu_chiral_2024}, viscoelastic cylindrical inserts can act as springs connecting the masses as shown in Figure~\ref{fig:UCMassSpring}. With this approach, we link the axial and torsional stiffness between the disks in a straightforward way to the geometry and material properties of the inserts, and automatically introduce viscous dissipation. For practical applications, this work presents the effect of viscoelasticity on the dispersion in an infinite CPC and on the vibration isolation efficiency of a finite CPC.

In a first step, a simplified viscoelastic material model assuming a constant storage modulus but frequency dependent loss modulus is used to analytically illustrate the role of damping on the interaction of multiple bandgaps in CPCs. The findings are then transposed to design physical CPCs with realistic viscoelastic inserts. To allow accurate numerical simulations, arbitrary frequency-dependent material properties are applied in all models. A fractional derivative material model allows for a systematic study of the material on the CPC's isolation efficiency. 
Finally, this approach is experimentally illustrated using three types of polyurethane foams with significantly different material properties. 

The remainder of the paper is structured as follows. After an analytical derivation of the role of damping in PCs, Section~\ref{Concept} shows the isolation efficiency of a finite size CPC and its dispersion. In Section~\ref{Tunability} the tunability of the crystal is explored, performing a parametric study on the viscoelastic model parameters by examining their effect on the dispersion and vibration isolation. Section~\ref{PracticalDesign} shows the models and manufacturing of a realistic design, and the experimental validation using three crystals with widely different material properties. Section~\ref{Conclusion} summarizes the main results of the paper. 

\section{Vibration isolation efficiency of chiral phononic crystals with viscous damping}\label{Concept}
In this section, the structural design of the chiral phononic crystal is described. The effect of frequency-dependent damping on bandgap generation in an analytical diatomic chain is analyzed. Then, the concepts of vibration isolation are presented in terms of force transmissibility and the dispersion calculation scheme that allows frequency-dependent materials is described. 

\subsection{Structural design of chiral crystals with viscoelastic inserts}

The design is based on the one-dimensional chiral phononic crystal described in \cite{bergamini_tacticity_2019,orta_inertial_2019}. Disks, the masses of the chain, are connected by tilted struts with spherical joint connections in a syndiotactic arrangement as shown in Figure~\ref{fig:UCMassSpring}. Between the disks, cylindrical viscoelastic inserts are glued. This design results in coupling between the rotation and translation of the disks, and thus leads to inertia amplification.
As shown in~\cite{ding_origin_2024}, the spherical joint connections do not contribute to the CPC's stiffness which now only depends on the viscoelastic insert geometry and the struts' inclination angle. The bandgap opening is thus defined by the axial and torsional stiffness of the inserts and can be adjusted by their shape and material properties.

In this work, the role of the viscoelastic material is investigated and the geometry of the crystal is kept constant as shown in Figure~\ref{fig:UCMassSpring}. The crystal disks are made of aluminium, have a diameter $D_{1}$ of 80~mm and a thickness $t_{1}$ of 7~mm.  The struts are made of steel with a length $L$ of 36~mm and diameter of 4~mm. The struts have an inclination angle $A$ of 45° relative to the disks. The viscoelastic inserts have a diameter $D_{2}$ of  40~mm and a thickness $t_{2}$ of 25~mm. The material properties are shown in Table~\ref{fig:Theoryproperties}. 

For the simplified analysis in this section, the storage modulus is assumed to be constant, using the low-frequency limit of the reference material presented in the parametric study of Sec.~\ref{Tunability}. The loss factor of the viscoelastic inserts is assumed to follow a theoretical frequency-dependent viscous damping relation defined by $\eta(\omega)=\frac{c\omega}{\beta}$ where $c$ represents the viscous damping coefficient and $\beta = 920$~kN/m the axial stiffness of one viscoelastic insert. 

\begin{table}[ht!]
\centering
\caption{Material properties.}
\begin{tabular}{|c|c|c|c|}
\hline
    & $E$ (MPa) & $\nu$& $\rho ~\mathrm{(kg/m^3)}$ \\
     \hline 
     Viscoelastic&17.70&0.2&1000 \\
     \hline 
     Aluminium& 71 000& 0.33 & 2700\\
     \hline 
     Steel & 200 000 & 0.33 & 7800\\
     \hline
\end{tabular}
\label{fig:Theoryproperties}
\end{table}

\subsection{Bandgap generation in diatomic phononic crystals with viscous damping}\label{sec:theory}
CPCs can be seen as equivalent diatomic phononic crystals \cite{ding_origin_2024}. This simpler model is first investigated in order to study the influence of damping on bandgap generation in CPCs. The archetypal phononic crystal is a periodic chain of masses and springs in which longitudinal waves propagate. In the theoretical case of identical masses $m$ and springs with stiffness $\beta$, a single bandgap opens at the masses' resonance frequency $\omega_0 = 2 \sqrt{\beta/m}$. Introducing two different masses $m_1$ and $m_2$, connected by springs with identical stiffness $\beta$, the additional degree of freedom leads to a more complex dispersion relation. According to~\cite{deymier_acoustic_2013}, the dispersion relation in a lossless diatomic chain can be calculated from the characteristic equation
\begin{equation}
    z^4 + z^2\left[2 - \left(2 - \frac{m_1\omega^2}{\beta}\right) \left(2 - \frac{m_2\omega^2}{\beta}\right)\right] + 1 = 0,
\end{equation}
where $z = \exp(ika)$ is the propagation constant, $a$ is the spacing between the masses and $k$ is the sought-after wave number. Two passbands, the so-called acoustical and optical bands, appear due to wave scattering by the system's internal dynamics. They are separated by a first bandgap between the masses' natural frequencies 
\begin{equation}
    \omega_{1,2} = \sqrt{\frac{2\beta}{m_{1,2}}}.
\end{equation}
A second bandgap opens at 
\begin{equation}
    \omega_3 = \sqrt{\omega_1^2 + \omega_2^2},
\end{equation}
which can be seen as the resonance of the equivalent mass $m_e = \frac{m_1 m_2}{m_1 + m_2}$.

Viscous damping can be introduced into this system in a straightforward manner, by setting the spring constant to the complex value
\begin{equation}
    \beta_v = \beta + i\omega c,
\end{equation}
in which $c$ is the viscous constant leading to a damping which increases linearly with frequency. Therefore, the second bandgap is affected differently from the first since the second resonance undergoes higher damping. 

\begin{figure}[ht!]
     \includegraphics[width=1\linewidth]{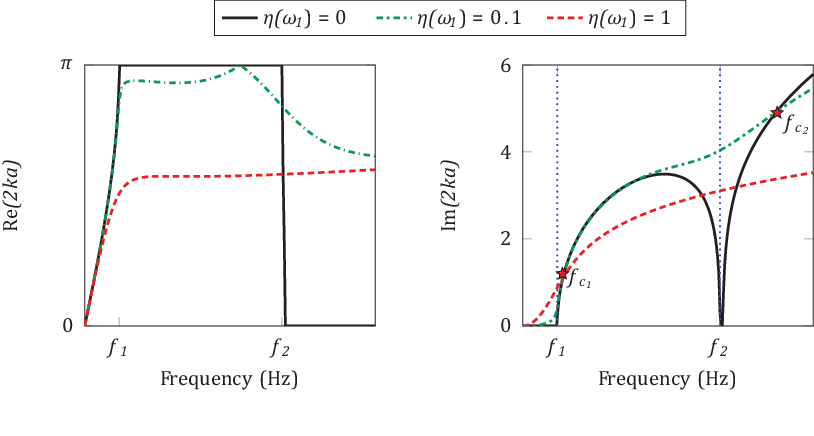} 
    \caption{Normalised real (left) and imaginary (right) part of the wavenumber for longitudinal waves as a function of frequency, plotted for the PC without damping ($\eta(\omega_1)=0$), with moderate damping ($\eta(\omega_1)=0.1$) and with high damping ($\eta(\omega_1)=1$). Dotted blue lines show the bandgap frequency limits $f_1$ and $f_2\approx f_3$. Red stars show the frequency limits before deterioration of the first bandgap $f_{c_{1}}$ and second bandgap $f_{c_{2}}$.}\label{fig:Dispersiontheory}
\end{figure}

Figure~\ref{fig:Dispersiontheory} illustrates the effect of the viscous damping on the bandgap formation by showing the normalised real and imaginary part of the wave number in a diatomic crystal. The mass  $m_1 = 0.006$~kg and $m_2 = 0.2$~kg are defined in order to match the acoustical and optical band frequencies of the CPC presented in the next section. The spring constant $\beta$ matches the viscoelastic insert's axial stiffness. In this case, the optical band is quite narrow, which is in line with the typical behaviour of a CPC. 

Two crystals with increasing viscous constants are compared to the lossless crystal. The viscous constant of the moderately damped CPC is defined so that the viscous loss factor equals 0.1 at the bandgap opening ($\eta(\omega_1)=0.1$) and the highly damped CPC defined so that the loss factor equals 1 ($\eta(\omega_1)=1$).  Without damping, the two bandgaps are sharply defined where $\imag(k) \neq 0$. Adding damping to the system introduces a smooth transition into the first bandgap, but considerably increases $\imag(k)$ in the optical band. For intermediate values of $\eta(\omega_1)$, the optical band can be further suppressed until both bandgaps are effectively merged. This does not affect $\imag(k)$ within the first bandgap, showing that the wave scattering effect is still sufficiently strong to suppress wave propagation. However, for higher values of  $\eta(\omega_1)$, internal resonances are overdamped and lower spatial attenuation is observed within the first bandgap. 

\begin{figure}[ht!]
    \centering \includegraphics[width=1\linewidth]{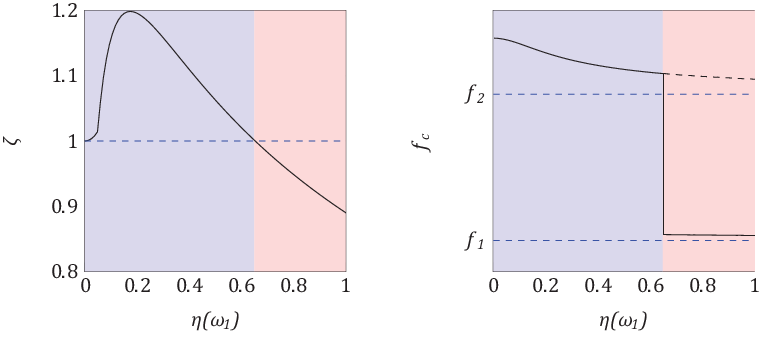} 
    \caption{(a) Relative efficiency of the bandgap in the presence of damping, $\zeta$, as a function of the loss factor at the first internal resonance $\eta(\omega_1)$. The red shaded area shows the overdamped regime and the blue area the underdamped regime. (b) Relation between the loss factor at the first internal resonance $\eta(\omega_1)$ and the frequency limit $f_c$ before bandgap's deterioration.}
    \label{fig:relationDampingEfficiency}
\end{figure}

In order to identify the transition between the intermediate and overdamped regime, the attenuation efficiency of the first bandgap is presented relative to the undamped case:
\begin{equation}
    \zeta = \frac{\text{max}\left( \imag(k)|_c \right)}{\text{max}\left(\imag(k)|_{c=0}\right)}.
\end{equation}
As shown in Figure~\ref{fig:relationDampingEfficiency}(a), for $\eta(\omega_1)$ values below $0.65$ $\zeta$ is suprisingly larger than 1: damping allows to merge the two bandgaps without deteriorating the first bandgap efficiency and therefore the maximum wave attenuation is more effective. However, deterioration of the second bandgap can be observed above a critical frequency $f_{c2}$. In the overdamped case, where $\zeta<1$, damping leads to a reduced isolation capacity even in the first bandgap above the critical frequency $f_{c1}$.

Figure~\ref{fig:relationDampingEfficiency}(b) shows the bandgaps' critical frequencies for bandgap deterioration. In the underdamped regime, the highest frequency before the isolation capacity deteriorates is displayed for the second bandgap, while in the overdamped regime attenuation deteriorates already in the first bandgap. For both cases, higher damping shifts the critical frequencies to lower values. 

This analytical model shows that viscous damping can suppress narrow pass bands, thereby effectively merging scattering bandgaps. Unlike locally resonant materials, the bandgap efficiency of phononic crystal is not substantially affected by damping up to a threshold damping value.  This fundamental result is the basis to investigate the role of viscoelastic materials in CPCs described in the remainder of this work.

\subsection{Isolation efficiency of chiral phononic crystal}
Numerical investigations on the realistic CPC's design described in Section~\ref{Concept}.1 are performed and compared to the analytical observations. The vibration isolation efficiency of the new design is described by analyzing wave propagation in the infinite periodic structure through dispersion curves. The identified bandgaps are compared to the dynamics of a finite structure by calculating the force transmissibility through a two unit cell crystal. 
\subsubsection{Dispersion analysis}
\begin{figure}[ht!]
    \centering
    \includegraphics[width=8cm]{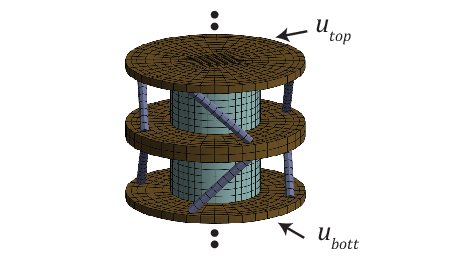}  
    \caption{ FE model of the meshed unit cell. Elements in light blue correspond to the viscoelastic inserts, in brown to the aluminium disks and in grey to the steel struts. }\label{fig:WFEM}
\end{figure}

The Wave Finite Element Method (WFEM) is used here to model the wave propagation in an infinite 1D periodic structure by applying Bloch-Floquet periodic boundary conditions on  a unit cell model finite element (FE) model \cite{mace_finite_2005,cool_guide_2024} shown in Figure~\ref{fig:WFEM}. The degrees of freedom at the bottom edge of the unit cell are related to the top edge through the modulation $u_{top }=u_{bott}e^{i2ka}$, which depends on the unknown Bloch wavenumber $k$ and length of the unit cell $2a$. The internal nodes are eliminated by dynamic condensation which leads to a simplified dynamic equation of motion. In this work, we will solve the eigenvalue problem using a direct approach to calculate the complex wavenumber $k$ as a function of the driving angular frequency $\omega$ and identify whether waves are propagating ($k$ is purely real) or spatially decaying ($k$ is complex) \cite{cool_guide_2024}. The calculation is done using an in-house code scripted in Matlab R2023b.

The unit cell FE model is built up in Ansys 2025R2. Disks and inserts are modeled using linear hexahedral solid elements while the tilted struts are discretized by beam elements to reduce the amount of degrees of freedom. The tips of the tilted struts are connected to the disks by spherical joints which constrain the relative displacement of the coincident nodes while rotation is permitted. No contact is defined between the inserts and the disks, but a conforming mesh is assigned in order to later constrain the degrees of freedom at their interface. The mesh presented in Figure~\ref{fig:WFEM} ensures convergence up to 3000~Hz, showing less than 5\% difference of the harmonic response results if it is further refined. The structural damping of the aluminium disks and steel struts is assumed as negligible, compared to insert's inherent damping and is therefore set to 0. The stiffness and mass matrices are exported to be used in the WFEM implementation. 

The stiffness matrix can be partitioned in nodes that correspond to the inserts and the ones corresponding to the disks and the tilted struts. 
To include frequency-dependent damping and stiffness, the stiffness matrix elements at the inserts' nodes are scaled for each driving frequency with complex values of the viscoelastic modulus defined in Equation~(\ref{complexModulus}). Therefore, the stiffness matrix becomes complex-valued and an additional damping matrix is not required. The conformal nodes at the interfaces between the inserts and the disks are finally constrained to ensure coupling between all parts.

\begin{figure}[ht!]
    \centering
    \includegraphics[width=1\linewidth]{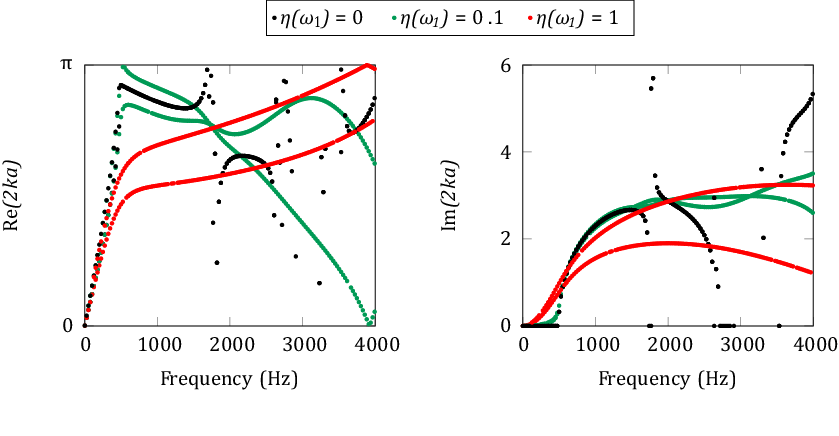} 
    \caption{Real (left) and imaginary (right) part of the wavenumber for longitudinal waves as a function of frequency, plotted for the CPC without damping ($\eta(\omega_1)=0$), with moderate damping ($\eta(\omega_1)=0.1$) and with high damping ($\eta(\omega_1)=1$). In dashed line is shown the threshold value for which the spatial attenuation per unit cell reaches 50\%.}\label{fig:DispersiontheoryCPC}
\end{figure}

The dispersion curves, which are showing an intricate combination of wave types for this structure, can be filtered by the polarization of the wave types, as defined in~\cite{bergamini_tacticity_2019,miniaci_spider_2016}. This allows for the identification of the longitudinal wave modes shown in Figure~\ref{fig:DispersiontheoryCPC}, which result from the translation and rotation coupling of the disks. Bending waves are less relevant in the scope of vibration isolation, and are therefore omitted. 

The calculated dispersion curves agree qualitatively with the theoretical model results in the previous section. 
In the presence of material damping, the start of the bandgaps is not uniquely defined. To allow a quantifiable and reproducible identification of the isolation frequency range for the parameter study in the next section, the frequency where the $\imag(k)$ reaches a threshold value of 10~m$^{-1}$ is chosen. For a unit cell size of $2a = 64$~mm, this is equivalent to 50\% wave attenuation per unit cell.

The CPC with moderate damping ($\eta(\omega_1)=0.1$) shows a bandgap starting at 500 Hz for both longitudinal waves. The observations made on the theoretical diatomic chain are validated by the practical design of the CPC. Increasing damping in the subcritical range does not negatively affect the imaginary part of the wavenumber while it suppresses internal resonances. However, in the case of overdamping ($\eta(\omega_1)=1$), a decrease of the imaginary part of the wavenumber is noticed and spatial attenuation is affected in the first bandgap.

\subsubsection{Harmonic analysis}
\begin{figure}[ht!]
    \centering
    \includegraphics[width=0.5\linewidth]{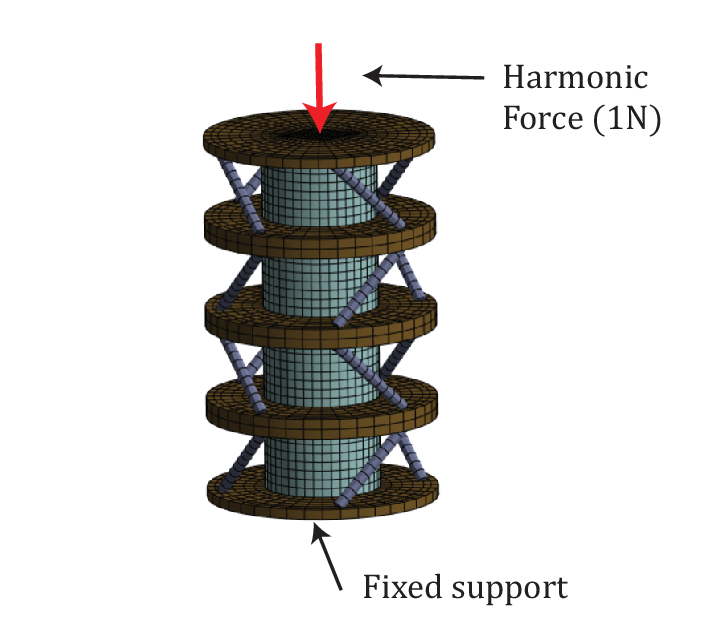} 
    \caption{FE model of the finite size crystal and its boundary conditions.}
    \label{fig:FEM Setup}
\end{figure}

The force transmissibility through a finite size chiral crystal is numerically calculated by building a two unit cell FE model in Ansys 2025R2 as shown in Figure~\ref{fig:FEM Setup}. The model definition of the finite crystal follows the description in the previous section. However, now a bonded contact region is defined between the inserts and the disks. The frequency-dependent viscoelastic properties are defined using the Ansys APDL command TB,PRONY~\cite{APDL_TB}. Using experimental tabular values, the software interpolates the viscoelastic properties at the requested frequencies using a Prony series. Additionally, boundary conditions and loading are defined as follows. The CPC is constrained at its bottom disk by a fixed support and is excited on the top face by a harmonic force of 1~N. The top disk is additionally constrained in its rotation to represent the attachment to a structure in real life applications. Using harmonic response analysis, the reaction force at the fixed support can be calculated and thus the force transmissibility can be determined. 

\begin{figure}[ht!]
        \centering
        \includegraphics[width=1\linewidth]{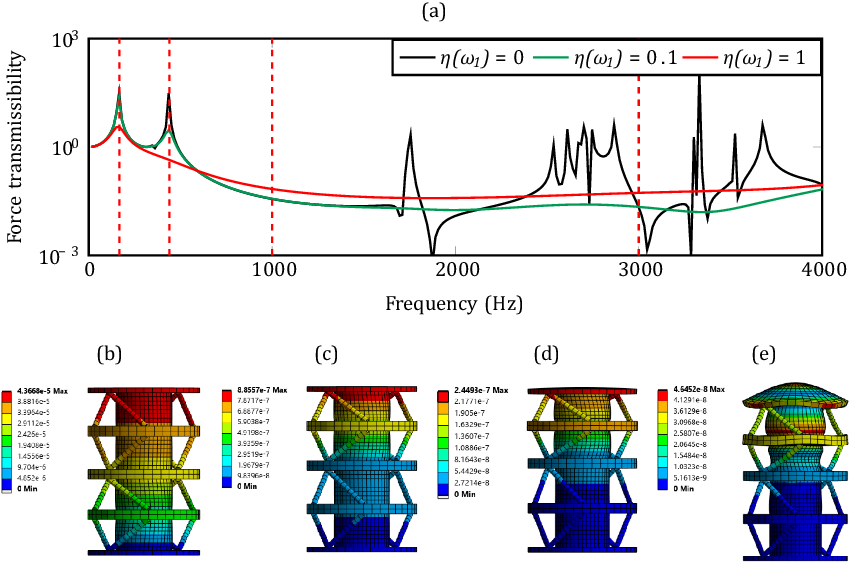} 
        \caption{(a) Force transmissibility of the two unit cell CPC with lossless ($\eta(\omega_1)=0$), moderately damped ($\eta(\omega_1)=0.1$) and highly damped ($\eta(\omega_1)=1$) viscoelastic inserts. Deformation of the two unit cells CPC at (b) 167Hz, (c) 440 Hz, (d) 1000 Hz and (e) 3000 Hz of the moderately damped CPC. The dashed lines represent the frequencies of the shown deformation patterns.}\label{fig:Deformation}
\end{figure}

The force transmission of the lossless CPC is compared to the damped CPC with moderate and high damping. The result is shown in Figure~\ref{fig:Deformation}(a). The CPC shows a clear isolation range from 480~Hz onwards. Increasing damping attenuates the low frequency resonance peaks, thereby broadening the attenuation frequency range. For the overdamped CPC ($\eta(\omega_1)=1$), the second resonance is entirely suppressed. Sharp resonance peaks above 1500~Hz are observed for the lossless CPC but are suppressed in the two damped CPC responses. However, the observation made on the dispersion relations correlates with the force transmissiblity of the finite CPC: overdamping decreases the attenuation level above 480 Hz. 

To illustrate the mechanism of the CPC, several deformation shapes for a damping value of $\eta(\omega_1)=0.1$ are shown in Figure~\ref{fig:Deformation}(b-e). Within the bandgap, e.g. at 1000~Hz, vibration is not transmitted to the bottom unit cell as observed in Figure~\ref{fig:Deformation}(d). The deformation of the crystal at resonance is shown in Figure~\ref{fig:Deformation}(b) and (c). 
In the high frequency range, above 2200~Hz, localized deformation modes of the struts and disks result in higher vibration transmission which is illustrated in Figure~\ref{fig:Deformation}(e).

\section{Tunability of chiral phononic crystal with realistic polyurethane materials}\label{Tunability}
This section explores the tunability of the new design using a wide range of typical frequency-dependent material properties of porous polyurethane materials for the inserts. To achieve this, a parametric study is performed and the effect of fractional derivative model parameters on the material master curves and the resulting vibration isolation is analyzed using dispersion curves and force transmissibility of the CPCs.

\subsection{Effect of fractional derivative model parameters on viscoelastic master curves}
The axial and torsional stiffness of the inserts can be tuned by  adapting their thickness, diameter and/or material properties. In this work, we investigate the influence of the inserts' material properties on the behavior of the CPC. The dimensions of the disks, struts and viscoelastic inserts, as well as the struts' inclination are detailed in Section \ref{Concept}. The inserts are made from polyurethane foam, a viscoelastic material widely used for vibration isolation due to its wide range of achievable properties. Polyurethane typically has a more complex frequency-dependency than what can be described by spring-dashpot arrangements such as the generalized Maxwell model~\cite{baz_active_2019,jalocha_revisiting_2015}.
As an alternative, fractional derivative models allow to capture the slope and variation of the frequency dependent storage and loss modulus  with a limited number of parameters \cite{lucie_rouleau_modelisation_2013}. In this work, the complex modulus of polyurethane materials is approximated by the four-parameters fractional derivative Zener model described in \cite{baz_active_2019} given by the relation 
\begin{equation}
    E(\omega)=\frac{E_{0}+(i\omega\tau)^\alpha E_{\infty}}{1+(i\omega\tau)^\alpha}
\end{equation}
with $E_{0}$ the (static) equilibrium modulus, $E_{\infty}$ the high frequency modulus of elasticity, $\alpha$ the order of the fractional derivative and $\tau$ the relaxation time.

\begin{figure}
    \centering
    \includegraphics[width=1\linewidth]{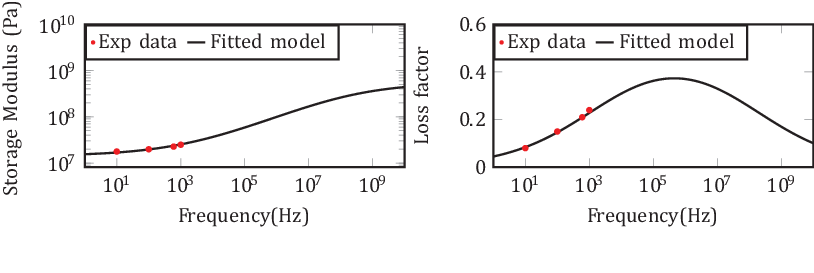} 
    \caption{Fit of the polyurethane material Sylodyn NF properties with a four parameter fractional derivative Zener model at room temperature. In red are shown extracted experimental material data from Getzner's documentation \cite{getzner_downloads_2026} and in black the fitted model.}
    \label{fig:Fit}
\end{figure}

A typical, moderately damped, porous polyurethane material from the company Getzner (Sylodyn NF \cite{getzner_downloads_2026}) is used as a reference. The four parameters of the fractional derivative model are fitted to the limited set of data points provided by the documentation using particle swarm optimization in MATLAB R2023. The fitted values are shown in Table~\ref{fig:FitModelParameters} and the fit between the data points and the fractional derivative model is shown in Figure~\ref{fig:Fit}. Although the product data do not ensure that the extrapolation to lower and higher frequencies is realistic, they cover most of the frequency range presented in this work. Therefore, the following study of the Zener model is only presented in this frequency band. An extension to higher frequencies would require the experimental determination of the master curves in a wider range, which can e.g. be done through dynamic mechanical analysis \cite{rouleau_characterization_2015}.

\begin{table}[ht!]
\centering
\caption{Fitted model parameters for the fractional Zener model applied on Sylodyn NF.}
\begin{tabular}{|c|c|c|c|}
\hline
     $E_{0}$(MPa)& $E_{\infty}$(MPa)&$\tau$(s) &$\alpha$\\
     \hline 
     14.2&550&1e-9&0.3 \\
     \hline 
\end{tabular}
\label{fig:FitModelParameters}
\end{table}

This material identification allows for a parametric study of the model parameters to illustrate the effect of the model parameters on the viscoelastic behavior of the material. To remain within a range of properties that is achievable with realistic materials, the model parameters are only moderately increased or decreased around the fitted values. It should be noted that in reality the parameters cannot be changed independently, so this study should be seen as a qualitative guideline relating material master curves to vibration isolation functionality. First, the equilibrium $E_{0}$ and high frequency modulus $E_{\infty}$ are both equally increased or decreased, changing the storage modulus but not the loss factor. In a second study, only the equilibrium modulus is scaled, without changing the high frequency modulus, which affects both the storage modulus master curve slope and the loss factor. 

\begin{figure}[ht!]
    \centering
\includegraphics[width=1\linewidth]{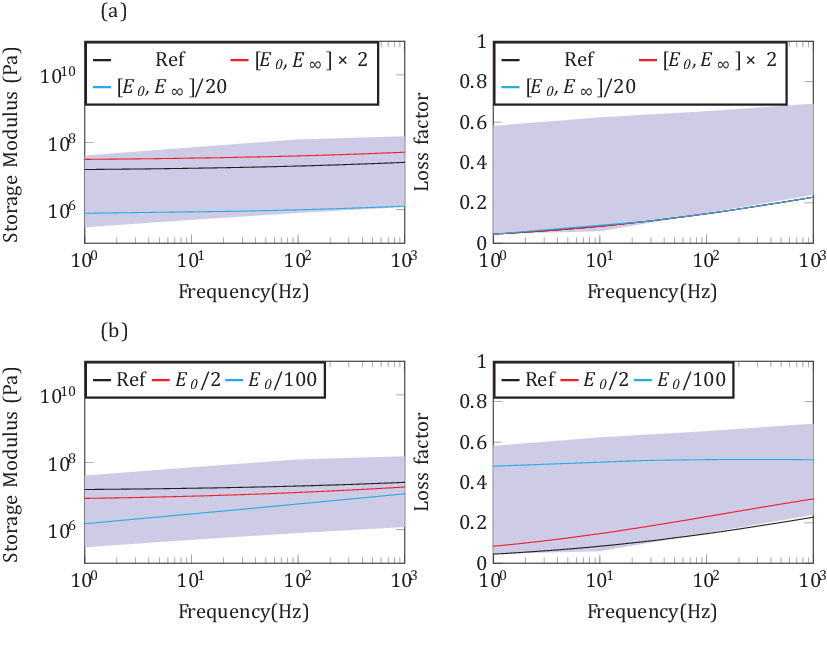} 
    \caption{Parametric study of the influence of the model parameters (a) $E_{0}$ and $E_{\infty}$ and (b) $E_{0}$ on the storage modulus (Pa) (left) and loss factor (right). The reference material is always presented in black. The shaded area represents the realistic range of material properties for polyurethane foams.}
    \label{fig:ParametricStudyMaterials}
\end{figure}

The resulting master curves of the storage modulus and loss factor for the parametric study are shown in Figure~\ref{fig:ParametricStudyMaterials}. The shaded areas represent a typical range of polyurethane material properties compiled from Getzner's documentation \cite{getzner_downloads_2026}. It can be observed that modifying model parameters has an impact on the storage modulus and loss factor within the shaded range representing a realistic range of material properties. Qualitative changes in the master curves include a simple linear rescaling, or a change in the slope of both stiffness and damping in the frequency range of interest. Depending on the parameter choice, constant damping or stiffness can be achieved, or, conversely, a strong frequency-dependency can be introduced. However, the common modeling assumption of constant stiffness and constant or linearly increasing damping is clearly a strong simplification which limits the accuracy for this family of materials.

\subsection{Influence of viscoelastic properties on the dispersion relation}
\begin{figure}[ht!]
    \centering
\includegraphics[width=1\linewidth]{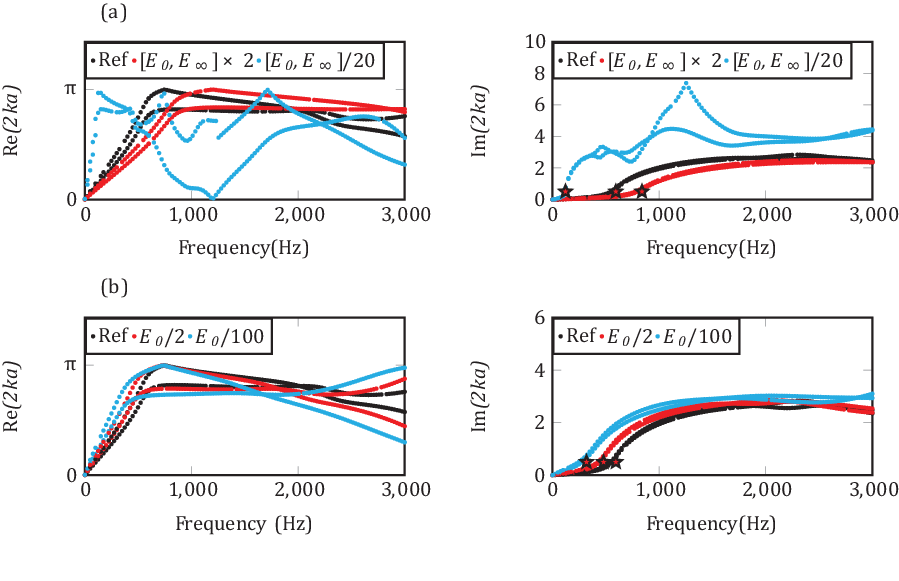} 
    \caption{Parametric study of the influence of the model parameters (a) $E_{0}$ and $E_{\infty}$ and (b)  $E_{0}$ on the real part (left) and imaginary part (right) of the wavenumber. The stars show the threshold level for the imaginary part of the wavenumber, indicating the start of the high-attenuation regime. In dashed line is shown the threshold value for which the spatial attenuation per unit cell reaches 50\%.}
    \label{fig:DispersionPolyurethane}
\end{figure}

Figure~\ref{fig:DispersionPolyurethane} shows complex dispersion curves for viscoelastic materials where only one model parameter is changed. Overall, the trend of the longitudinal wave dispersion -- responsible for force transmission -- does not change. In all cases, two longitudinal wave branches are visible, with distinct real wave numbers but very similar imaginary wave numbers, meaning that both waves decay equally fast. The start of the effective attenuation range, denoted by a star, clearly correlates with the stiffness, which can be seen in all four cases. As expected in PCs, softer materials lead to a lower frequency for the start of the bandgap. In the extreme case of the blue curve in Figure~\ref{fig:DispersionPolyurethane}(a), the very soft core yields a much more irregular dispersion curve compared to all other cases. Internal dynamic deformations of the soft inserts are the cause for this, but this effect does not deteriorate the imaginary part of the wavenumber. It can also be seen that all curves of the imaginary part of the wavenumber reach similar maximum values between 35 and 40~m$^{-1}$. In general, the softer materials lead to slightly higher maximum values. Once again, the softest material in Figure~\ref{fig:DispersionPolyurethane}(a) is an outlier reaching even higher values. The relation between stiffness and vibration isolation efficiency will be quantified in the next section.

The transition from pure viscous damping to bandgap attenuation, which happens where the slope of the imaginary wave number changes, is influenced by damping. In Figure~\ref{fig:DispersionPolyurethane}(b), the black curve shows a more sudden change in slope than the blue and red lines, although it happens at a higher frequency (and thus, at higher stiffness) than the blue line. In this case, the blue curve corresponds with a material with much higher loss factor.

\subsection{Influence of viscoelastic material properties on the force transmissibility}
\begin{figure}[ht!]
    \centering
    \includegraphics[width=1\linewidth]{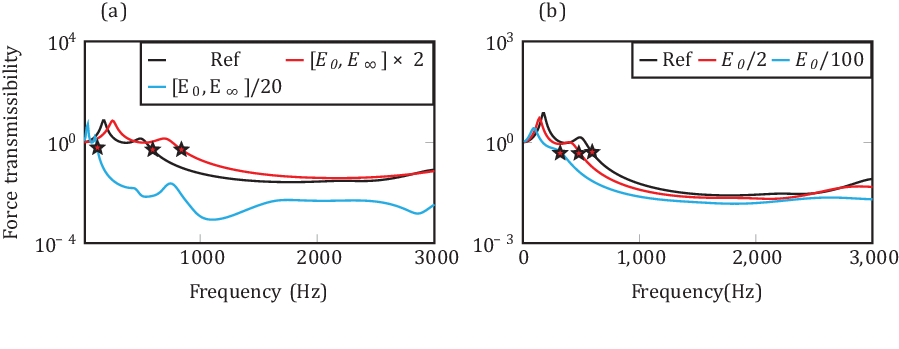} 
    \caption{Parametric study of the influence of the model parameters (a) $E_{0}$ and $E_{\infty}$ and (b) $E_{0}$ on the force transmissibility of the two unit cells chiral crystal. The stars indicate the start of the high-attenuation regime, at the frequencies retrieved from the dispersion curves in Figure~\ref{fig:DispersionPolyurethane}.}
    \label{fig:ParametricIL}
\end{figure}

The force transmissibility of the finite crystal, consisting of only two unit cells, is calculated for the same set of material parameters following the methodology defined in Section~\ref{Concept}. The results of the parametric studies are shown in Figure~\ref{fig:ParametricIL} and reflect the conclusions of the dispersion relations. If the storage modulus varies significantly, the overall stiffness of the crystal shifts the isolation range. Additionally, the softer crystals yield a clearly lower force transmissibility. However, this comes at the practical cost of structural stability: for practical purposes the support stiffness cannot be decreased arbitrarily. The material damping, which changes the onset of the imaginary wave number as described in the previous section, here influences the amplitude of the resonance at the start of the isolation range. Looking at the cases in Figure~\ref{fig:ParametricIL}(b), it can be seen that the blue and red curves, which show the highest damping, show the lowest resonant amplitudes irrespective of the frequency where they occur.

The parametric studies provide insights into the tunability of the crystal. Adjusting the equilibrium $E_{0}$ and high frequency $E_{\infty}$ modulus together allows to shift the attenuated frequency range without affecting the amplitude of the resonant peaks. Shifting the force transmissibility to lower frequency while reducing the amplitude of the resonant peaks can be done by reducing the equilibrium modulus $E_{0}$ but this could be problematic for structural integrity reasons. 

\begin{figure}[ht!]
    \centering

    \includegraphics[width=1\linewidth]{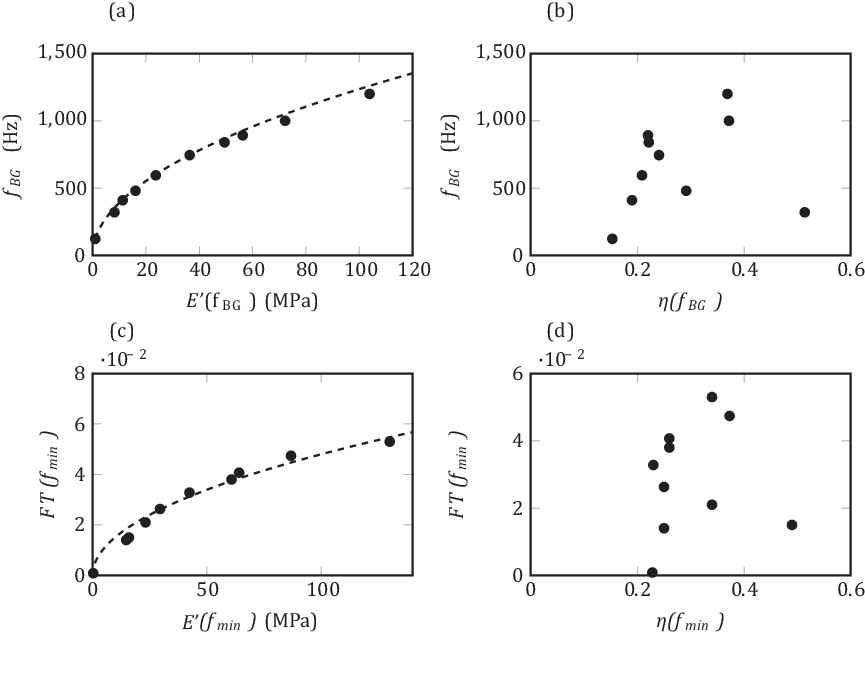} 
    \caption{Correlation studies between isolation metrics and material properties. Bandgap opening frequency $f_{BG}$ and storage modulus evaluated at the bandgap opening $E'(f_{BG})$ (a), bandgap opening frequency $f_{BG}$ and loss factor evaluated at the bandgap opening $E'(f_{BG})$ (b), minimum force transmission $FT(f_{min})$ and storage modulus evaluated at  the minimum's frequency $E'(f_{min})$ (c) and minimum force transmission $FT(f_{min})$ and loss factor evaluated at  the minimum's frequency $\eta(f_{min})$ (d). The fitted power laws relating the storage modulus and the opening bandgap frequency and the minimum force transmission are displayed in dashed lines.}
    \label{fig:Correlations}
\end{figure}

Figure~\ref{fig:Correlations} shows a quantitative study on the tunability of the CPCs by investigating two important metrics: the starting frequency $f_{BG}$ of the bandgap (denoted by stars in Figures~\ref{fig:DispersionPolyurethane} and \ref{fig:ParametricIL}) and the minimum of the force transmission curve $FT(f_{min})$, which is a measure for the isolation efficiency. They are plotted against the storage modulus and loss factor, representing the material properties at the relevant frequencies $f_{BG}$ and $f_{min}$ (frequency at minimum force transmission). In this study, a broader range of material properties are considered by modifying the model parameters $E_0$, $E_\infty$,  $\tau$ and $\alpha$ independently. The chosen parameters respect the range of properties that can be achieved by realistic polyurethane foams displayed in Figure~\ref{fig:ParametricStudyMaterials}. All the properties of the CPCs used are described in Table~\ref{tab:PropertiesCPC}. 

The results show a clear correlation between the storage modulus and the two isolation metrics. The simulation results are fitted using a mean square method to a power law function using Matlab r2023b curve fitting toolbox. Figure~\ref{fig:Correlations}(a) shows a fitted trend $f_{BG} \sim E'^{1/2}$, which corroborates that the bandgap indeed opens at a resonance frequency governed by the square root of the insert stiffness. Furthermore, the power law trend between the minimum transmitted force and the bandgap opening frequency observed in Figure~\ref{fig:Correlations}(c) shows that an increase in stiffness leads to higher force transmission, and thus a less effective bandgap.

Figure~\ref{fig:Correlations}(b,d) present the relation between the isolation metrics and the loss factor. In this case, no trend can be distinguished. Moreover, the results show that a low $f_{BG}$ and a low force transmissibility can be reached for the highest loss factors. This is advantageous, since the resonance amplitude at the opening of the bandgap can be reduced through higher damping. Therefore, this quantitative study shows that the tuning of the vibration isolation of a CPC can be unrelated to material damping.

\begin{table}[ht!]
    \centering
    \caption{Material model parameters for the quantitative study.}
    \begin{tabular}{|c|c|c|c|c|c|c|c|c|c|c|}
      \hline
    CPC n° & 1& 2& 3& 4& 5& 6& 7& 8& 9& 10\\
    \hline
    \hline
         $E_0$(MPa)&14.2& 28.4&7.1&  0.7&  14.2&  14.2&  14.2&  14.2&7.1&0.14\\
         \hline
         $E_{inf}$(MPa)& 550&1100&275&27.5&  550&  550&  550&  550&  550&  550\\
         \hline
         $\tau$(s)&$1e^{-9}$&$1e^{-9}$&$1e^{-9}$&$1e^{-9}$&$1e^{-9}$&$1e^{-9}$&$1e^{-6}$&$2.5e^{-7}$&$1e^{-9}$&$1e^{-9}$\\
         \hline
         $\alpha$&0.3&0.3&0.3&0.3&0.2&0.25&0.3&0.3&0.3&0.3\\
         \hline
    \end{tabular}
    
    \label{tab:PropertiesCPC}
\end{table}

\section{Practical design and experimental validation}\label{PracticalDesign}
In this section, an experimental validation of the previous observations is performed. This requires adequate manufacturing of CPCs using a variety of polyurethane foams to illustrate the accuracy of the models and prove the tunability functionality of the soft inserts.

\subsection{Manufactured design}
\begin{figure}[ht!]
   \centering
    
    \includegraphics[width=1\linewidth]{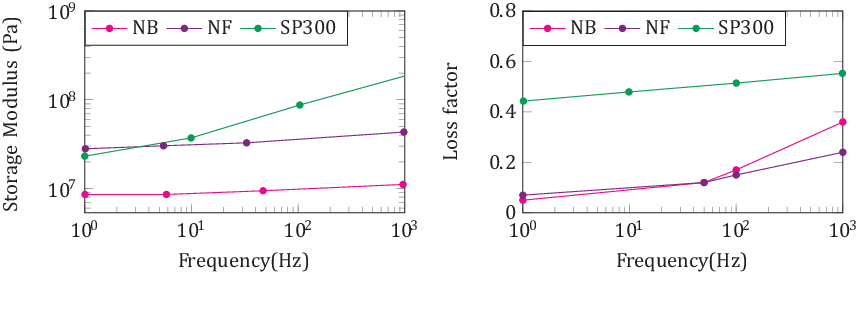} 
    \caption{Storage modulus (left) and loss factor with respect to frequency (Hz) of the materials NB (pink), NF (purple) and SP300 (green). Extracted experimental material data from Getzner's documentation \cite{getzner_downloads_2026}.}
    \label{fig:ExpMasterCurve}
\end{figure}

Three CPCs with different soft inserts are manufactured. To validate the previous observations on the effect of material properties on the force transmissibility, the materials are selected such that two of them have a comparable storage modulus but different loss factors, and two have a comparable loss factor but different storage modulus.  The three viscoelastic materials chosen for the comparison are the materials Sylodyn NB (soft) and NF (stiff) and the material Sylodamp SP300 (high-damping) manufactured by Getzner \cite{getzner_downloads_2026}. The materials NF and NB show a comparable loss factor but different storage modulus: NB's storage modulus is 20 times lower than NF's loss factor. The materials NF and SP300 show a storage modulus in the same order of magnitude, however the slope coefficient of the storage modulus of SP300 differs significantly from the slope of material NF. Furthermore, material SP300 shows a much higher loss factor than material NB and NF. Their properties are shown in Figure~\ref{fig:ExpMasterCurve}. The dimensions of the disks, struts and inserts remain the same as previously defined for the numerical parametric study. 

\begin{figure}[ht!]
    \centering
    
    \includegraphics[width=1\linewidth]{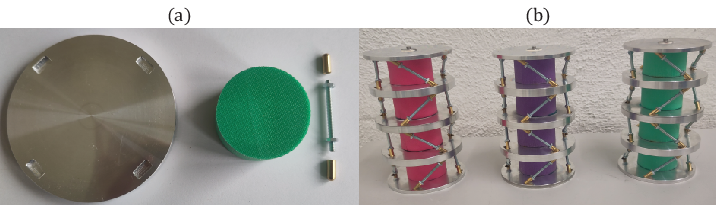} 
    \caption{(a) Picture of the assembly components: on the left the disk, in the middle the insert and on the right the strut. (b) Picture of the three manufactured crystals: on the left the crystal with material NB, in the middle the crystal with material NF and on the right the crystal with material SP300.}
    \label{fig:Pieces}
\end{figure}

The crystals are manufactured in three parts as shown in Figure~\ref{fig:Pieces}(a). The disks are manufactured with four slits milled on both sides. The cylindrical inserts are waterjet cut from a material sheet with the desired thickness. The struts consist of threaded rods on which brass spherical cap nuts are screwed. The inserts are glued to the disks and the struts are slotted into the disks. The contact between the struts and disks ensures a spherical connection between them: at the contact point the struts can rotate but no relative displacement is possible due to the compression between the disks provided by the inserts. The assembly of the three crystals is shown in Figure~\ref{fig:Pieces}(b).

\subsection{Experimental set up}
\begin{figure}[ht!]
    \centering
    \includegraphics[width=0.5\linewidth]{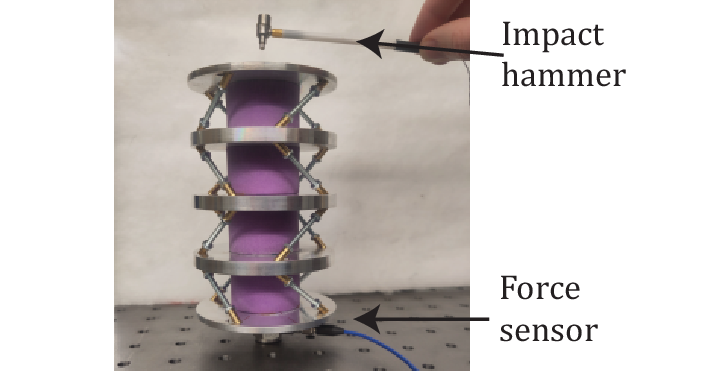} 
    \caption{Picture of the experimental set up with the crystal with material NF.}
    \label{fig:ExpSetup}
\end{figure}

The force transmissibility of the three crystals is measured using the experimental set-up shown in Figure~\ref{fig:ExpSetup}. The bottom disk of the crystal is screwed on a force sensor (PCB 208C01). The crystal is excited by an instrumented hammer on its top disks. The input force is measured by the internal sensor of the 2~g hammer (PCB 086E80). The signals are captured using an in-house coded Labview software interface connected to a NI PXI-4496 data acquisition card. The force transmissibility is calculated taking the average of the Fourier transform of ten impulse responses.

\subsection{Experimental validation of the numerical models}
The force transmissibility of the three manufactured crystals is compared to an FE model to investigate the predictability of the bandgap location and their efficiency. The storage modulus master curves retrieved from the data sheets are rescaled to fit the experimental force transmissibility, to account for the sample's shape factor and unknown shear modulus information. Details on the fitting procedure are given in Appendix~\ref{sec:Appendix}. 

\begin{figure}[ht!]
   \centering
    
    \includegraphics[width=1\linewidth]{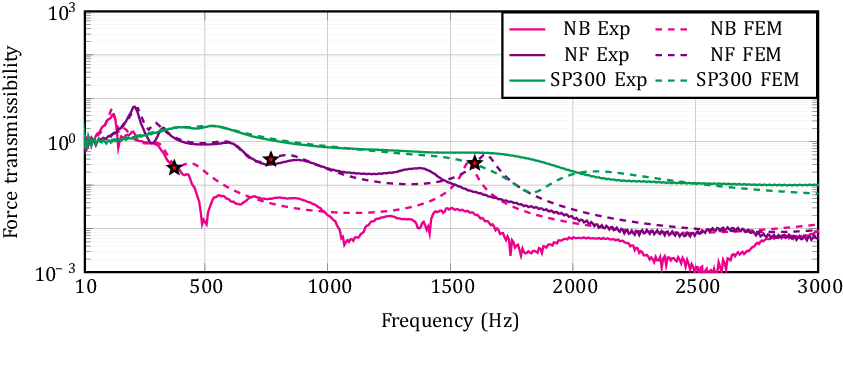} 
    \caption{Force transmissibility measured (continuous line) and numerically calculated (dashed line) of the crystals with material NB (pink), material NF (purple) and SP300 (green).}
    \label{fig:TFExp}
\end{figure}

\begin{figure}[ht!]
   \centering
    \includegraphics[width=1\linewidth]{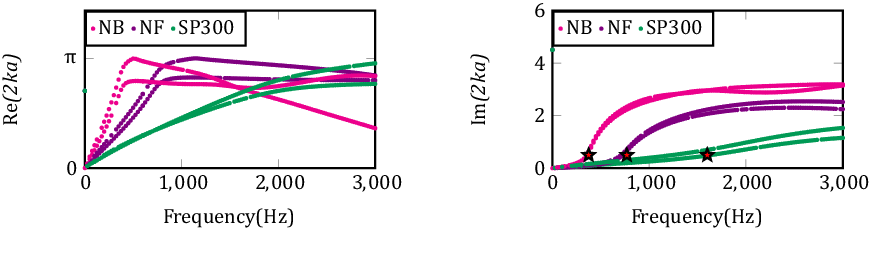} 
    \caption{Real (left) and imaginary (right) part of the wavenumber with respect to frequency (Hz) of the CPC's with material NB (pink), NF (purple) and SP300 (green).}
    \label{fig:DispersionExp}
\end{figure}

The three crystals allow to validate the observations made in the previous numerical investigations. Lower storage moduli shift the force transmissibility to lower frequencies as observed in Figure~\ref{fig:TFExp} between the CPCs with NB and NF inserts. The longitudinal bandgap opening for the crystal with inserts NB is observed in the dispersion curves at lower frequency than for the one with stiffer inserts NF as shown in Figure~\ref{fig:DispersionExp}. Increasing the slope  of the storage modulus and the loss factor decreases the attenuation of the CPC with the inserts SP300. From the dispersion curve of the crystal with inserts SP300, it can be observed that the longitudinal bandgap opens at a frequency higher than 1500~Hz, which is explained by the much higher storage modulus in this frequency range. The high loss factor makes it almost impossible to distinguish between the viscous damping regime and the bandgap. However, the structural resonant peak in the force transmissibility of the CPC with these inserts is significantly attenuated due to the higher loss factor. Overall, the experimental and numerical results for force transmissibility agree well. Stronger deviations can be seen for material NB. This can be explained by experimental difficulties for such a soft material, leading to imperfections in the setup. The study for these three materials does confirm the conclusions of the parametric study: the isolation frequency range and efficiency is governed by the storage modulus, whereas the loss factor has an impact on the resonant amplitudes.

\section{Conclusion}\label{Conclusion}

In this paper, a practical design has been proposed to improve the tunability and broaden vibration isolation of CPCs by means of viscoelastic damping. The stiffness and damping of the spring elements connecting the masses of the crystal are governed by the material properties and shape of cylindrical viscoelastic inserts. Glued between the disks, they undergo translation and rotational deformation due to hinged tilted connectors. The results show that, for a wide range of realistic loss factors, damping does not deteriorate the vibration isolation capacity of the crystal but rather merges multiple distinct bandgaps into one wide attenuation range. Using WFEM dispersion calculations and FE harmonic response analysis, a design procedure has been detailed to estimate the vibration isolation efficiency of this new design. The particular case of strongly frequency-dependent and highly damping polyurethane is presented.

The tunability of CPCs for vibration isolation is illustrated by a parametric study based on the Zener fractional derivative model. The material parameters are varied to show that the storage modulus correlates with the start of the isolation frequency range, and with the maximum force reduction. The loss factor, while at subcritical values, on the other hand, does not correlate with the isolation efficiency, since the bandgap mechanism is stronger than the viscous damping effect on the longitudinal wave propagation. For this particular type of CPC, the damping can thus be used to reduce resonant effects, whereas the stiffness defines the isolation properties. These observations are consistent with an analytical model of a diatomic chain, which shows that increasing viscous damping in the underdamped regime reduces internal resonances without affecting the spatial attenuation within the bandgaps up to a threshold value. 

The real-life applicability of these findings was validated experimentally by three CPCs with different insert properties. The experimental results validated the numerical observations: increasing the loss factor in the full frequency range reduces the amplitude of the resonant peaks while increasing the storage modulus shifts the attenuated frequency range to higher frequencies and decreases the vibration reduction efficiency. Good agreement is observed between the numerical and experimental force transmissibility curves after rescaling of the storage modulus to account for the unknown shear modulus. 

This study shows that this class of CPCs can be tuned to meet the criteria for real-life applications: static stiffness, isolation frequency range and required force transmission reduction. The tuning can be performed by varying material properties, but also by changing the insert geometry, the disk size, and the strut inclination. The work can be extended to other strongly viscoelastic materials such as rubber. The main practical caveat is the uncertainty of the frequency-dependent material master curves, for which data sheets typically offer only limited information. Detailed material testing such as dynamic mechanical analysis might be necessary to cover the frequency range of interest with high enough accuracy to allow for predictive models.

\section*{Acknowledgements}
This work was done in the framework of the European Commission's Horizon Europe research and innovation programme under grant agreement No 101072415 and funded by the Swiss State Secretariat for Education, Research and Innovation (SERI), contract No 22.00225. Views and opinions expressed are however those of the authors only and do not necessarily reflect those of the European Union. The  European Union cannot be held responsible for them. Internal Funds KU Leuven are gratefully acknowledged for their support.

\appendix
\counterwithin{figure}{section}
\section{Appendix}\label{sec:Appendix}

\begin{figure}[ht!]
    \centering
    \includegraphics[width=1\linewidth]{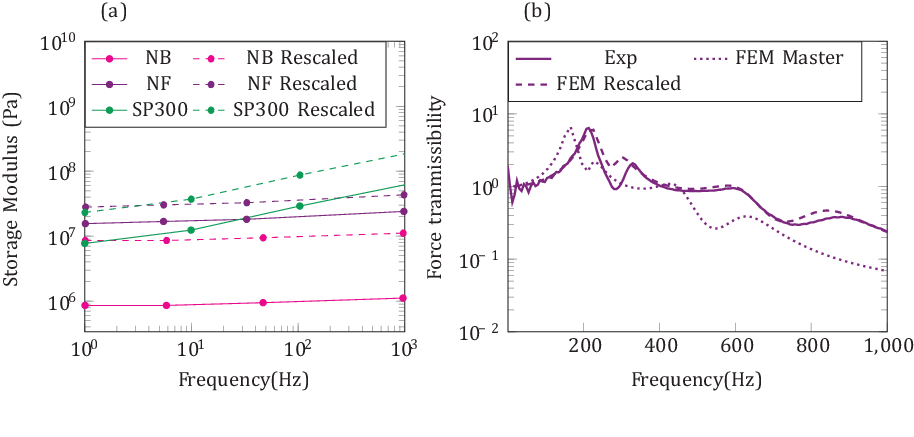}  
    \caption{(a) Master curves of the material NB, NF, SP300 and rescaled master curves to match experimental results. (b) Experimental curves (continous), numerically calculated with material properties following the master curves (dotted) and numerically calculated with rescaling of the master curves (dashed lines) of the force transmissibility for the CPC with material NF.}
    \label{fig:ComparisonRescaling}
\end{figure}

In this study, the material definition of the viscoelastic inserts is assumed linear and isotropic to allow numerical calculations in Ansys. This assumption, with a typical Poisson ratio of 0.2, overestimates the ratio between the dynamic shear and storage modulus compared to the data provided by the documentation. For instance, for material NF, the dynamic shear modulus specified by the documentation is about 1.48 MPa while the shear modulus assuming a linear isotropic model, a dynamic modulus of 14.94 MPa and a poisson ratio of 0.2 is about 6.22 MPa.   For pure compression, the Young's modulus master curve, before rescaling, is sufficiently accurate, whereas for torsion, a good model of the shear modulus is equally necessary. Comparing the experimental to the numerical force transmissibility using the data sheet master curves, the numerical curve is similar in shape and amplitude but shifted to a lower frequency as shown in Figure~\ref{fig:ComparisonRescaling}. The storage modulus is therefore linearly rescaled until the first resonance of the numerical model matches the first resonance of the experimental curves. This scaling preserves the frequency-dependent trends, but corrects the storage modulus to match reality. The loss factor is not rescaled and follows the documentation's master curve. The correction requires about a multiplication factor for the storage modulus of 10 for the inserts NB, of 3 for the inserts SP300 and of 1.8 for the inserts NF. The softest material requires the largest scaling, which can be explained by a larger relative influence of the shear modulus. 

\label{sec:Bibliography}
%--------------------------------------------------%
\bibliographystyle{unsrtnat}
\bibliography{bib/ViscoPaper}
\end{document}